\documentclass[a4,journal]{IEEEtran}

\usepackage{amsmath,amsfonts}
\usepackage{algorithmic}
\usepackage{algorithm}
\usepackage{array}
\usepackage[caption=false,font=normalsize,labelfont=sf,textfont=sf]{subfig}
\usepackage{textcomp}
\usepackage{stfloats}
\usepackage{url}
\usepackage{verbatim}
\usepackage{graphicx}
\usepackage{cite}

\usepackage{siunitx}
\usepackage{placeins}
\usepackage[Option]{overpic}
\usepackage{tabularx,booktabs}
\usepackage{multirow}
\usepackage{xcolor}
\usepackage[capitalise]{cleveref}
\begin{document}

\title{Self-sustaining Ultra-wideband Positioning System for Event-driven Indoor Localization}

\author{Philipp Mayer,~\IEEEmembership{Graduate Student,~IEEE,}
        Michele Magno,~\IEEEmembership{Senior Member,~IEEE,}\\
        Luca Benini,~\IEEEmembership{Fellow,~IEEE,}


	\thanks{
		Manuscript received XXXX XX, 2022; revised XXXX XX, 2022; accepted
        XXXX XX, 2022. Date of publication XXXX XX, 2022; date of current version XXXX XX, 2022. \textit{(Corresponding author: Philipp Mayer.)}
		
        Philipp Mayer is with the Integrated Systems Laboratory, ETH Zurich, 8092 Zurich, Switzerland (e-mail: mayerph@iis.ee.ethz.ch).
        
        Michele Magno is with the Center for Project-Based Learning, ETH Zurich, 8092 Zurich, Switzerland (e-mail: michele.magno@pbl.ee.ethz.ch).
        
        Luca Benini is with the Integrated Systems Laboratory, ETH Zurich, 8092 Zurich, Switzerland, and also with the Department of Electrical, Electronic and Information Engineering, University of Bologna, 40136 Bologna, Italy (e-mail: luca.benini@unibo.it).
	}
}

\markboth{}%
{Shell \MakeLowercase{\textit{et al.}}: A Sample Article Using IEEEtran.cls for IEEE Journals}


\maketitle

\begin{abstract}
Smart and unobtrusive mobile sensor nodes that accurately track their own position have the potential to augment data collection with location-based functions. To attain this vision of unobtrusiveness, the sensor nodes must have a compact form factor and operate over long periods without battery recharging or replacement. This paper presents a self-sustaining and accurate ultra-wideband-based indoor location system with conservative infrastructure overhead. An event-driven sensing approach allows for balancing the limited energy harvested in indoor conditions with the power consumption of ultra-wideband transceivers. The presented tag-centralized concept, which combines heterogeneous system design with embedded processing, minimizes idle consumption without sacrificing functionality. Despite modest infrastructure requirements, high localization accuracy is achieved with error-correcting double-sided two-way ranging and embedded optimal multilateration.
Experimental results demonstrate the benefits of the proposed system: the node achieves a quiescent current of \SI{47}{\nano\ampere} and operates at \SI{1.2}{\micro\ampere} while performing energy harvesting and motion detection. The energy consumption for position updates, with an accuracy of \SI{40}{\centi\metre} (2D) in realistic non-line-of-sight conditions, is \SI{10.84}{\milli\joule}. In an asset tracking case study within a \SI{200}{\metre\squared} multi-room office space, the achieved accuracy level allows for identifying 36 different desk and storage locations with an accuracy of over \SI{95}{\percent}. The system’s long-time self-sustainability has been analyzed over \SI{700}{days} in multiple indoor lighting situations.
\end{abstract}


\section{Introduction}
\IEEEPARstart{I}{nternet} of Things (IoT) sensors are becoming smart and unobtrusive, surrounding us in everyday life. With the drastic increasing number of low-cost and connected sensing systems, there is also emerging demand for pervasive location services \cite{Ahmed2020}. Next to the prominent examples of navigation and asset tracking \cite{Zhao20, Elsanhoury22}, the location awareness of sensors can significantly widen the scope of mobile IoT systems. Location-awareness not only allows to assign events of interest to their spatial origin but also enables location-based services where the sensor adapts and interacts with the surrounding \cite{Cheema18, LiAnna22}.

Commercial solutions based on the global navigation satellite system (GNSS) allow accurate localization at a moderate power consumption of several \si{\milli\watt} in always-on operation for outdoor applications. In contrast to that, there is no widely deployed solution for precise localization in indoor and hybrid scenarios. This is particularly interesting as the market potential for indoor location services is expected to be significantly larger compared to the outdoor counterpart \cite{Cheema18}. Thus, there has been significant research effort in technologies allowing GNSS-like indoor localization functionality in recent years.

Focusing on device-based systems with high localization accuracy, IEEE 802.15.4z ultra-wideband (UWB) is more and more becoming the gold standard as it promises robustness against multipath and shadowing phenomena \cite{Jeon21,Coppens22,Flueratoru2021}. By opting for UWB, recent research works achieve localization accuracies in the range of a few tens of centimeters and below within highly idealized environments with perfectly aligned antennas. If dynamic real-world indoor scenarios with non-line-of-sight (NLOS) propagation are considered, localization accuracy decreases significantly \cite{Stefano2010, Silva2020, Barbieri21}. Advanced postprocessing \cite{Kegen19, Yang22, Vlad2022}, and optimized transmission schemes \cite{Tiemann2019,Laadung22} allow to mitigate the drastic accuracy loss, however, at the cost of energy consumption and scalability \cite{Tiemann2019, Grosswindhager2019, Corbalan2019}.

One major drawback of today’s commercial UWB transceivers is their peak power consumption of a few hundred milliwatts, which implies significant stress on the design of small-sized battery-powered devices \cite{Jimenez2016, Mayer2019-WFIOT}. This drawback can be circumvented, at the cost of reduced functionality, by outsourcing energy-demanding signal processing to the infrastructure \cite{Costanzo2018, Pannuto2018, Fabbri20}. If these limitations in terms of bi-directionality and range are not acceptable, there inevitably arises the need to recharge the battery of the UWB-enabled device periodically. Unfortunately, the requirement to actively recharge reduces the device’s long-term reliability and, in general, prevents a pervasive operation \cite{Vankecke15}.

Next to the peak consumption, which is critical for the storage-element selection, it is also crucial to reduce the idle consumption of any wireless sensor node to allow long-time operation. For UWB-enabled devices, this is particularly challenging, as in contrast to the transmitter, which can be implemented energy-efficiently \cite{Jazairli10, Allebes21}, the energy cost for reception is significantly harder to curtail and tends to dominate. Thus, it is vital that no periodical synchronization between the battery-powered device and infrastructure is required. In addition, recent trends in the design of smart-sensing devices with ”near-sensor” processing and emerging wireless communication technologies such as Narrowband Internet of Things (NB-IoT) and Long-Range (LoRa) impose additional demanding requirements on the sensor’s power supply and energy management.

A mature technology to prolong sensor lifetime without manual battery replacement or recharge is the use of environmental energy sources with energy harvesting (EH) \cite{Tuncel2021}. However, making EH suitable for small-sized IoT devices requires advanced design techniques to compensate for the intermittent and low power nature of EH transducers \cite{Adu-Manu2018, Newell2019}. This includes the conditioning of the typical low-voltage and low-power output signal of the EH transducer and the optimization of the harvesting efficiency with maximal power point tracking. In addition, there is the need to temporal match the fluctuating environmental energy with the load requirements. Finally, there is not only the necessity to study and optimize the circuitry itself, but also to precisely analyze the application-specific energy intake and power consumption statistics to ensure long-time availability.

\begin{figure}[!t]
\centering
\includegraphics[width=0.48\textwidth]{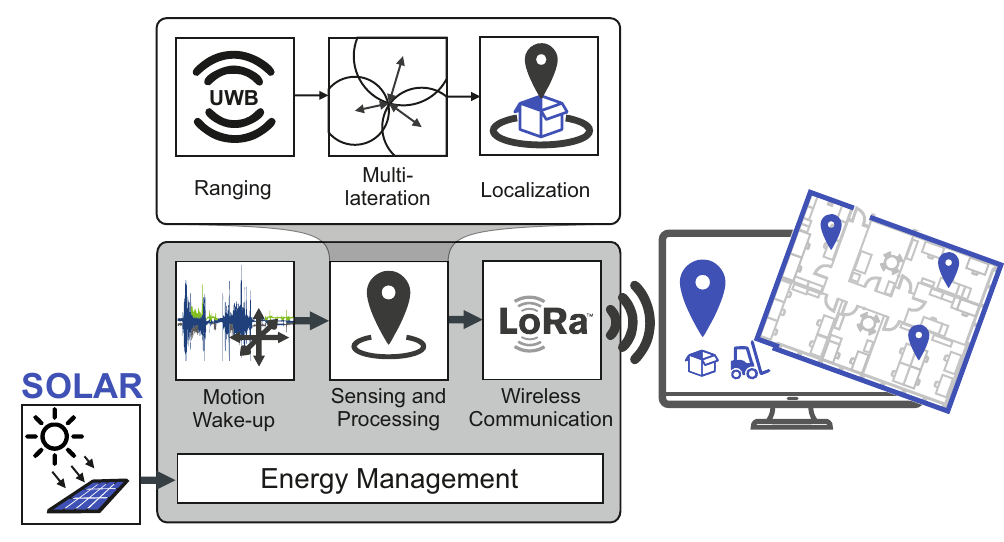}
\caption{High-level overview of the proposed self-sustainable asset tracking system based on motion-triggered ultra-wideband (UWB) distance measurements and embedded processing.}
\label{fig:motivation}
\end{figure}

This work presents the design of an EH-powered indoor asset tracking system, visualized in \cref{fig:motivation}, that balances the limited environmental energy with the energy requirements of state-of-the-art UWB and LoRa transceivers for location and communication, respectively. In the proposed architecture, this is achieved by a tag centralized event-driven operation, where the tag remains in an ultra-low power sleep mode most of the time and autonomously initiates the communication with the mains-supplied localization infrastructure. The combination of a motion-triggered wake-up circuitry with on-board processing allows for a significant reduction of energy-demanding message exchange. Self-sustainability, despite low-light environmental conditions, is achieved by the presented heterogeneous system design optimized for nano-quiescent current operation. 

In particular, this paper presents four main contributions:

\begin{enumerate}
\item{A hardware-software co-designed tag based on the event-driven sensing paradigm that enables the spatial-temporal limiting of energy consumption. The proposed design allows a quiescent consumption of \SI{4.68}{\micro\watt} while enabling motion-triggered localization at \SI{40}{\centi\metre} accuracy.}
\item{The design and evaluation of a self-sustainable indoor localization system that combines embedded tag centralized processing, event-driven sensing, and solar energy harvesting.}
\item{The full implementation of the system combined with an in-depth analysis of the long-time self-sustainability by combining a model-based design approach with \SI{700}{\day} real-world indoor lighting dataset.}
\item{A real-life application scenario evaluation of the full system and in realistic non-line-of-sight (NLOS) conditions demonstrating accurate asset tracking in a \SI{200}{\metre\squared} sized multi-room office environment.}
\end{enumerate}

The rest of the article is organized as follows: 
Section \ref{section:relatedWork} presents the recent literature and discusses the proposed work in its context;
Section \ref{section:background} introduces the principles and implemented algorithms for time-of-flight (ToF) indoor localization;
Section \ref{section:architecture} describes the proposed system architecture for energy-neutral event-driven localization; 
Section \ref{section:results_EH} characterizes and models the circuits energy consumption and validates the self-sustainability; 
Section \ref{section:results_localization} reports experimental results in the targeted application of indoor asset tracking;
Finally, section \ref{section:conclusion} concludes the paper.

\section{Related Work}
\label{section:relatedWork}
There has been significant research effort in technologies allowing accurate indoor localization. This effort resulted in a multitude of methods spanning from vision- and RF-based approaches \cite{Zafari2019} to the analysis of light, pressure, vibration, or the electromagnetic field \cite{Maheepala2020, Alam2021}. A comparison of highly topical works on device-based indoor localization exploiting different underlying technologies is given in \cref{table:comparison}. 

\begin{table*}
\caption{Comparison to state-of-art works targeting device-based indoor localization.}
\centering
\renewcommand{\arraystretch}{1.35}
\begin{tabular}{@{}l*{7}{c}c@{}}
\toprule
 & Sensors'20 \cite{Arbula20} & TIM'21\cite{Kusche2021} & SEN'22 \cite{Ye22} & IOT'21 \cite{Flueratoru2021} & RFID'20 \cite{Dardari20}  & \bfseries This Work \\ 
\midrule
Localization technology  & infrared & magnetic field & BLE & UWB (3db Access)  & UWB (custom) & UWB (DW3000) \\ 
Energy harvesting & no & no & no & no &  wireless &  solar \\ 

\multirow{2}{*}{Localization error} & \multirow{2}{*}{\SI{40}{\centi\metre} (2D)} & \multirow{2}{*}{\SI{1}{\metre} (2D)} & \multirow{2}{*}{\SI{59}{\centi\metre} (2D)} &\SI{10}{\centi\metre} (2D)   & \multirow{2}{*}{\SI{4}{\centi\metre} (2D)}  &  \SI{40}{\centi\metre} (2D)\\ 
&  &  &  & \SI{37}{\centi\metre} (3D)  &   & \SI{1.2}{\metre} (3D) \\

Sampling rate & \SI{1}{\hertz} & $<$\SI{100}{\hertz} & $<$\SI{50}{\hertz} &   $<$\SI{500}{\hertz} & \SI{5}{\hertz} \scriptsize{$^a$}  & \bfseries event-driven \\


Energy per localization & - & \SI{120}{\micro\joule} & \SI{22}{\micro\joule}  & \SI{140}{\micro\joule} &  \bfseries \SI{1}{\micro\joule}  & \SI{10.84}{\milli\joule} \scriptsize{$^b$} \\

Range & \SI{20}{\metre}  & \SI{8}{\metre} & \SI{5}{\metre} &  \bfseries over \SI{100}{\metre} & \SI{22}{\metre} \scriptsize{$^a$}   & \bfseries over \SI{100}{\metre} \\

Tested area & \SI{30}{\metre\squared} (LOS) & \SI{100}{\metre\squared} (LOS) & \SI{92}{\metre\squared} (LOS) & \SI{16}{\metre\squared} (LOS)& \SI{70}{\metre\squared} (LOS)  &  \bfseries \SI{200}{\metre\squared}  (NLOS) \\

Processing scheme  & offline & offline &  offline & offline & offline  & \bfseries online \\

Energy neutrality & no & no & no  &  no & \bfseries yes & \bfseries yes \\

Deployment effort & high & high & high  & \bfseries low  & high & \bfseries low \\

\bottomrule
\end{tabular}
\scriptsize{\\ \vspace{3px} \hspace{250px} $^a$ Limited by EH subsystem. \hspace{20px} $^b$ Full system consumption.}\\
\label{table:comparison}
\end{table*}

Infrared-based systems such as the one presented in \cite{Arbula20} promise energy efficiency and low device cost but are limited to line-of-sight applications. Kusche et al. \cite{Kusche2021} presented a novel approach based on an artificial magnetic field. Although this concept could be implemented energy-efficiently, robustness and accuracy are limited. State-of-the-art Bluetooth low energy (BLE) with angle of arrival (AoA) functionality allows reaching 2D localization errors close to a half meter with an energy consumption of only \SI{22}{\micro\joule} \cite{Ye22}. Similar to the UWB-based work presented in \cite{Dardari20}, this is achieved by outsourcing most of the complexity to the infrastructure. If NLOS applications with accuracies of \SI{50}{\centi\metre} and below are required, UWB is the most promising technology. Although analyzing accuracy and energy consumption only in LOS or quasi LOS configuration, the results of \cite{Flueratoru2021} and \cite{Dardari20} demonstrated impressive performance for UWB. In \cite{Flueratoru2021}, the UWB IC from \textsc{Qorvo} (DW1000) and \textsc{3db Access} (3DB6830) were analyzed. The commercial ICs provide high configuration flexibility allowing a best-case accuracy of \SI{10}{\centi\metre} with sample rates of up to \SI{500}{\hertz} and an energy consumption per localization down to \SI{140}{\micro\joule}. Finally, Dardari et al. \cite{Dardari20} combined a custom UWB pulse generator with an ultra-high frequency (UHF) wake-up radio and wireless energy harvesting. In the tested optimal environmental conditions, this resulted in an energy-neutral tag with cold-start capability that reached a localization accuracy of a few centimeters. It is important to notice that the reported numbers are a proof-of-concept based on a few static positions with high oversampling. Beyond that, recent research presented in \cite{Allebes21} shows significant potential to improve energy efficiency in today's UWB transceivers.

In contrast to previous works, the focus of the presented asset tracking system is self-sustainability and high localization accuracy in a non-idealistic real-world scenario. This second objective requires a more power-hungry UWB configuration and transmission scheme, resulting in a significantly higher energy consumption per localization. The challenging opposing requirements are addressed by combining various algorithmic and architectural design techniques. By opting for novel tag centralized processing combined with event-driven localization, the presented approach allows for significantly reduced system active time without compromising latency. The paper further shows the benefits of a heterogeneous hardware architecture combined with hardware-tailored power management to minimize the system's quiescent currents to \SI{1.2}{\micro\ampere} despite active motion wake-up circuitry. The combination limits the energy consumption to the absolute necessary functional block, only activating energy-demanding subsystems when required. By extending the system with solar energy harvesting, the presented solution achieves self-sustainable operation despite consuming \SI{10.84}{\milli\joule} per localization due to the robust ranging, on-board processing, and long-range communication. This work demonstrates energy autonomy by applying a model-based design approach combined with longtime environmental data. Finally, the full implementation in a credit card-sized shape is analyzed in a real-world multi-room office environment demonstrating a 2D accuracy below \SI{50}{\centi\metre}.

\section{Background}
\label{section:background}
In contrast to conventional radio transmission using modulated carrier waves to encode information, UWB is based on short and consequently high bandwidth pulses. This property allows time-of-flight measurements with sub-nanosecond accuracy \cite{ghavami2007}. The ToF information between devices can be used to estimate distances and ultimately for localization by applying multiliteration algorithms.

\subsection{DS-TWR algorithm} 
\label{sub:doubleSidedTW}

\begin{figure}[t!]
    \centering
    \includegraphics[width=0.45\textwidth]{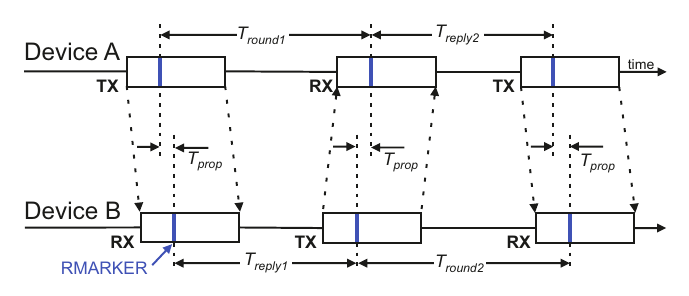}
    \caption{Double-sided two-way ranging with three messages between two devices.}
    \label{fig:dstwr}
\end{figure}

A simple solution for ToF measurements between two unsynchronized devices is the concept of two-way ranging (TWR). TWR uses time-stamped messages between bidirectional communicating devices to determine the round-trip time. Subsequently, the average propagation time can be calculated from the timing information. The limited frequency stability of the oscillators in both devices when using TWR causes a ranging error which increases with the distance between the nodes. To suppress this error while keeping the computation overhead low double-sided two-way ranging (DS-TWR) with three messages, shown in \cref{fig:dstwr}, can be applied. The algorithm uses two round trip measurements where the reply of the first round trip is used as the initiator of the second one. The average signal propagation time ($T_{prop}$) can be calculated according to \cref{equation:dstwr}.
\begin{equation}
T_{prop} = \frac{T_{round1}T_{round2}-T_{reply1}T_{reply2}}{T_{round1}+T_{round2}+T_{reply1}+T_{reply2}}
\label{equation:dstwr}
\end{equation}
Where the round trip times $T_{round[1,2]}$ are defined as the time from transmitting a message to receiving the response, and the reply times $T_{reply[1,2]}$  represent the time from receiving a package to sending.

Applied on the tag centralized indoor localization approach, the tag (battery-powered and mobile) initiates the DS-TWR algorithm by sending a time-stamped broadcast message to all anchors (mains-powered and fixed) in the transmission range. Subsequently, the anchors add a receive and transmit timestamp to the message and respond in the order of their ID. The timing information from the previous round-trip messages is used to apply double-sided ranging. Thus, a second broadcast is required in deep duty-cycled operation, doubling the energy consumption of DS-TWR. If two round-trip and replay times to every anchor node are determined, propagation times and corresponding distances can be calculated.

A detailed analysis of clock drifts and their influence on ranging errors for different ToF methods is given in \cite{Tschirschnitz19}.

\begin{figure}[t!]
    \centering
    \includegraphics[width=0.4\textwidth]{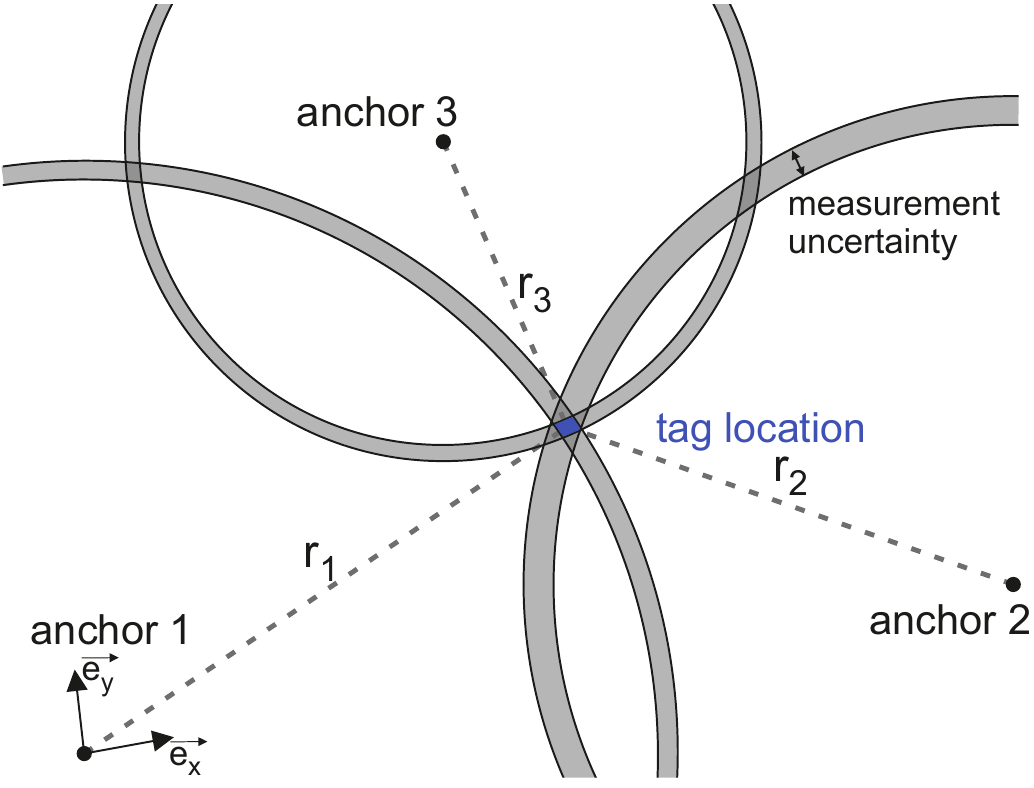}
    \caption{Multilateration to determine the tag position based on range measurements to known anchor points in 2D.}
    \label{fig:trilateration}
\end{figure}

\subsection{Multilateration} 
\label{sub:trilateration}
When the distance to at least three anchors with a known position is measured, the tag's absolute position can be calculated with multilateration. Multilateration algorithms calculate the position under consideration that the point of interest has to lie on the intersection point of spheres with radii similar to the range measurements centered at the anchor locations, \cref{fig:trilateration}. The underlying range measurements’ accuracies are crucial to precisely calculate the point of interest as inaccuracies inevitably result in a larger volume in which the tag must be located.

An effective and sufficiently lightweight multilateration algorithm that allows the implementation on performance and memory constraint microcontrollers (MCU) is presented in \cite{Larsson2019}. The algorithm, which has been implemented for this work on a \SI{32}{bit} \textsc{ARM Cortex-M4} operating at \SI{80}{\mega\hertz}, assumes that the measurement errors can be modeled as Gaussian noise with zero means. In such a case, multilateration can be described as a non-convex optimization problem. With the approach presented in \cite{Larsson2019}, the non-convex optimization problem can be simplified to an eigenvalue problem, solvable non-recursive and non-iterative fashion with a guaranteed global optimality.

\section{System Architecture}
\label{section:architecture}
\begin{figure}[!t]
    \centering
    \includegraphics[width=0.5\textwidth]{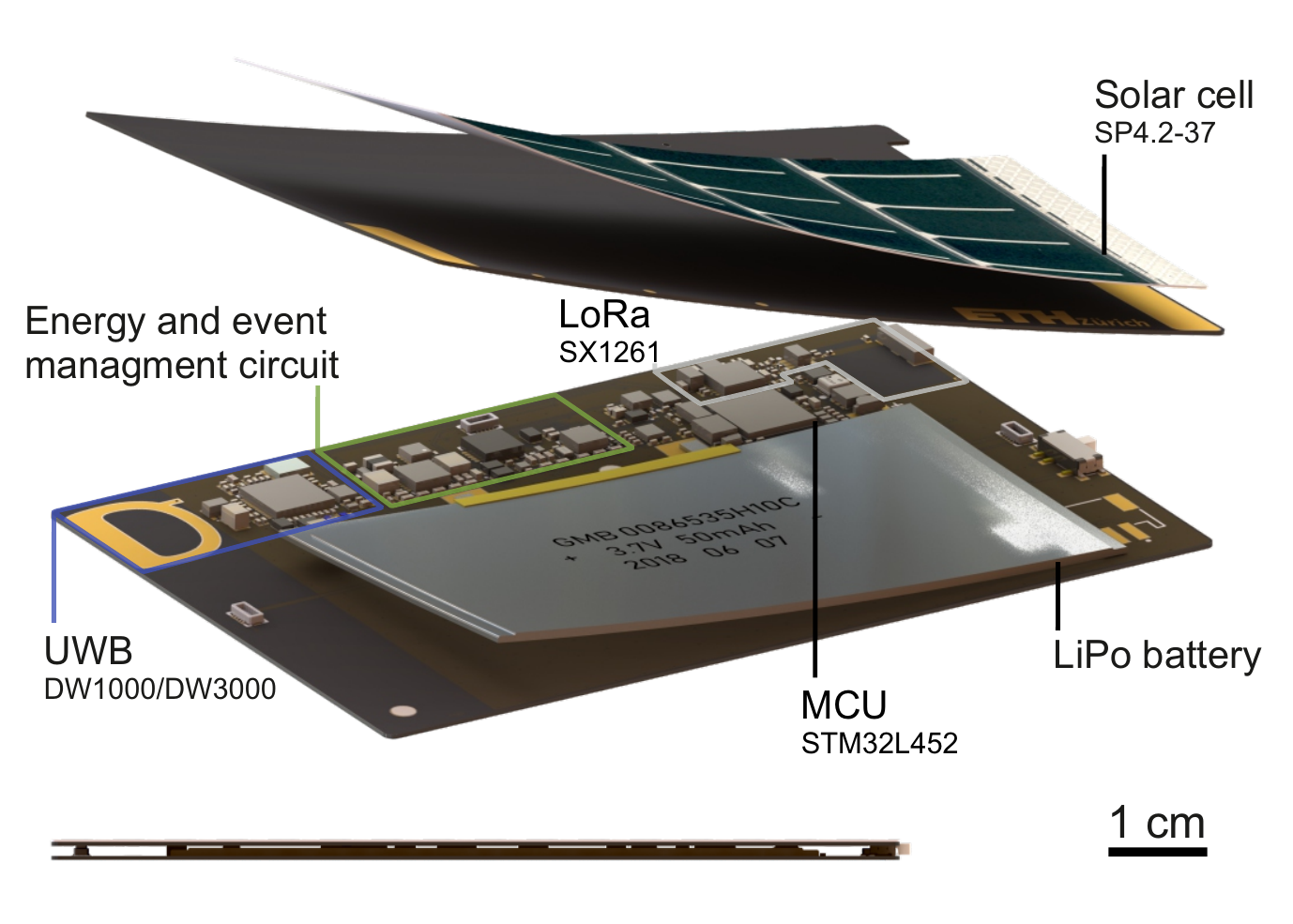}
    \caption{Prototype of the designed self-sustainable indoor localization tag in the shape of an \SI{85}{\milli\metre} $\times$ \SI{55}{\milli\metre} $\times$ \SI{2}{\milli\metre} business card.}
    \label{fig:randering}
\end{figure}

\cref{fig:randering} shows the developed and implemented credit card-sized tag, comprised of a solar cell-based energy harvesting subsystem and an application-specific smart sensing circuit. The system exploits an event-driven sensing scheme combined with tag-centralized processing to balance the consumed energy with the limited harvested energy. To ensure a high quality of service despite the fluctuating nature of the ambient energy, source and load power points are decoupled with a rechargeable battery.

The following subsection \ref{subsection:circuit-overview} gives an overview of the hardware architecture. Subsection \ref{subsection:designConsiderations} discusses the implemented design steps to achieve self-sustainable operation. Finally, subsection \ref{subsection:localization} describes the event-driven tag-centralized localization approach.

\subsection{Circuit Overview}
\label{subsection:circuit-overview}
The circuit is built of two main functional domains, a highly efficient energy- and event management circuit and a power-hungry localization circuit based on UWB, as shown in  the top part of \cref{fig:blockdiagram}. This strict division enables system-level power-gating and thus low leakage currents in deep duty-cycled or event-driven operations.

\vspace{3px}
\subsubsection{Energy- and event management circuit}
\label{subsubsection:circuit-power}
\begin{figure}[!t]
    \centering
    \includegraphics[width=0.48\textwidth]{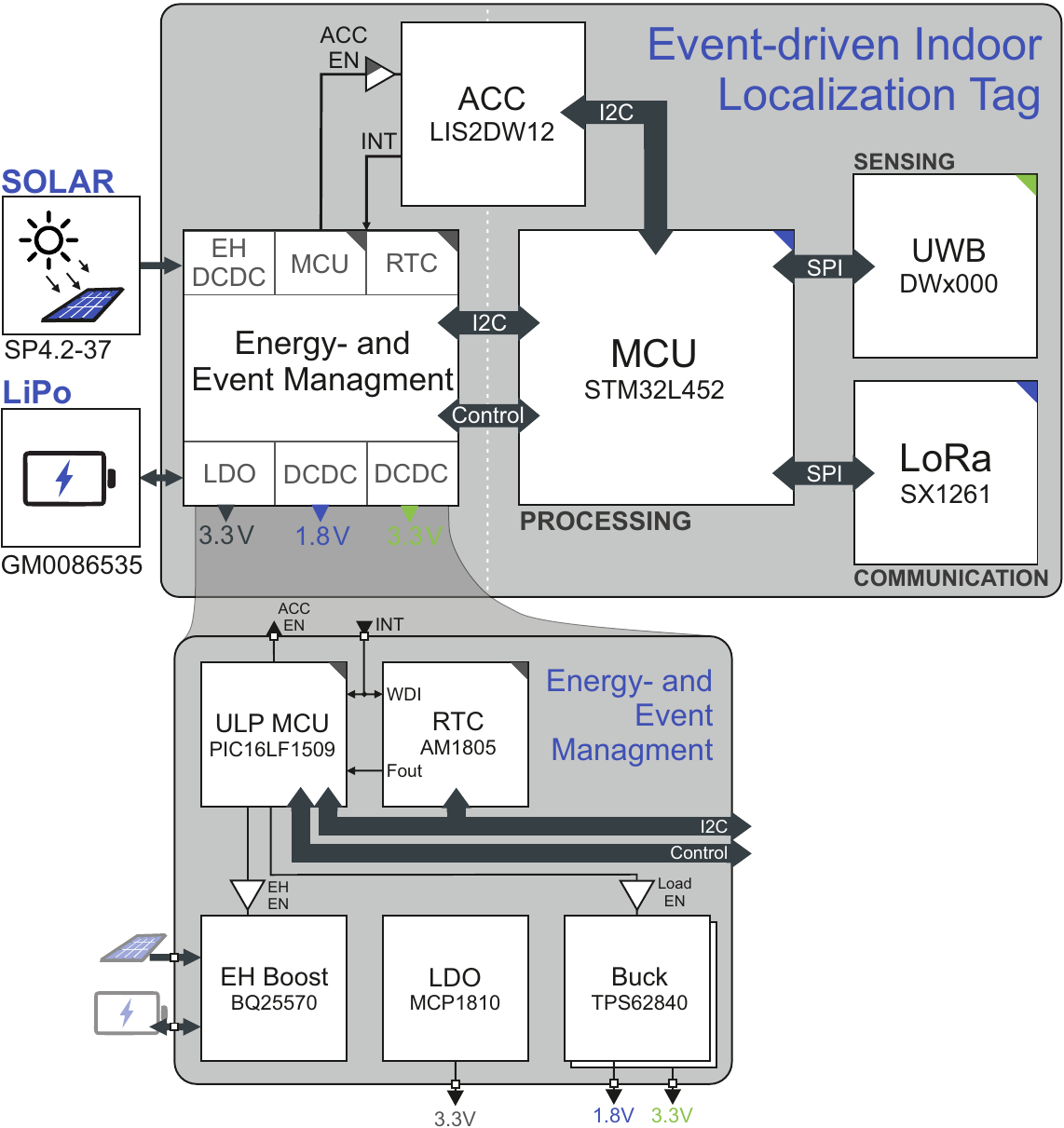}
    \caption{High-level block diagram of the proposed embedded system. A highly efficient always-on domain performs energy and event management, while the power-hungry sensing domain is only activated during event-driven localization events.}
    \label{fig:blockdiagram}
\end{figure}

The firmware configurable energy- and event management circuit, highlighted in the lower part of \cref{fig:blockdiagram}, is based on the design shown in \cite{mayerSPU}. Its primary purpose is the reduction of the system's quiescent consumption alongside decreasing the active time of power-hungry system blocks. In addition, the circuit aims at efficiently using the limited available energy by supplying the sensor node with high conversion efficiency in all power modes from sub--\si{\micro\watt} sleep up to hundreds of \si{\milli\watt} during localization. This is achieved by a programmable and configurable circuit that allows the power-gating of system blocks when their function is not essential in the operating sequence.

The central building block is a programmable 8-bit PIC16LF1509 microcontroller with 14kB flash memory. Its main purpose is to increase the system's energy efficiency by configuring the sensor node in favorable functional modes and executing lightweight processing tasks such as wake-up prefiltering. The MCU, which runs at \SI{31}{\kilo\hertz} in active mode,  consumes \SI{20}{\nano\ampere} and \SI{4.8}{\micro\ampere} in memory retentive sleep mode and active mode, respectively. An AM1805 sub-threshold real-time clock (RTC) further reduces the system's quiescent consumption by offloading timekeeping features to a dedicated highly efficient (\SI{11}{\nano\ampere}) application-specific integrated circuit (ASIC).

The energy harvesting functionality is based on a BQ25570 boost converter that achieves high harvesting efficiency due to integrated maximal power point tracking (MPPT) for a wide input current range, thus well suited for varying light conditions in indoor scenarios. As an energy harvesting transducer, a flexible solar cell SP4.2-37 from \textsc{PowerFilm} with an active area of $\SI{70}{\milli\metre} \times \SI{37}{\milli\metre}$ is used. The cell is selected with a typical operating voltage of \SI{4.2}{\volt}, which allows a low boost converter ratio and results in efficient voltage conditioning. The source and load power points are decoupled with a \SI{50}{\milli\ampere\hour} pouch cell lithium polymer (LiPo) battery. The supply of the individual switchable power domains, indicated with colored triangles in \cref{fig:blockdiagram}, is done with dynamic load optimized TPS62840 buck converters. In addition to the switching regulators, the circuit hosts the MCP1810 IC, a nano quiescent current optimized low dropout regulator (LDO). Its quiescent current of  \SI{11}{nA} enables an efficient supply of the energy- and event management circuit's always-on domain.

Event-driven operation is enabled by a LIS2DW12 accelerometer configured in motion wake-up mode. The configurable sensitivity and sampling rate allow operation with an average current consumption down to \SI{380}{\nano\ampere}.

\subsubsection{Localization circuit}
\label{subsubsection:circuit-localization}
The localization circuit is based on the concept of a “smart sensor” comprising functional blocks for data acquisition  \textit{SENSING}, near sensor computation  \textit{PROCESSING}, and data exchange  \textit{COMMUNICATION}.

\textit{SENSING} - The backbone of the indoor position functionality is UWB based on the wireless transceiver DW3000 from \textsc{Qorvo} (formally \textsc{Decawave}). To allow maximal range and thus a low number of anchors per area, channel 5 (\SI{6489.6}{\mega\hertz}) is used with a channel bandwidth of \SI{499.2}{\mega\hertz} and a mean equivalent isotropically radiated TX power (EIRP) of \SI{-41.3}{dBm/MHz}. Furthermore, the data rate has been limited to \SI[per-mode = symbol]{850}{\kilo bit\per\second}, and the preamble is configured to \SI{2048}{symbols}. 

\textit{PROCESSING} - During active sensing, the tag's functionality is orchestrated by an STM32L452 microcontroller (single-core \SI{32}{bit} \textsc{ARM Cortex-M4} with floating point unit) operating at \SI{80}{\mega\hertz}. Next to the sequence control during a localization event, the microcontroller is used to initialize the subsystems, execute the double-sided two-way ranging, and calculate the tag position with embedded multiliteration. 

\textit{COMMUNICATION} - Independently of UWB, the system comprises the \textsc{Semtech} SX1261 transceiver, a second wireless interface for communication with data-collection infrastructure via LoRa. The receiver is configured to operate at a spreading factor SF7 with a bandwidth of \SI{250}{\kilo\hertz}. Although the payload of the UWB would allow using the transceiver not only for ranging but also for data communication, opting for LoRa has the distinct advantages of wider range and easy integration in existing widely available IoT infrastructure. This enables applications in campus-area networks or sensing tasks besides pure localization, e.g., workplace exposure monitoring \cite{Sherazi18,Wu21}.

\subsection{Energy-efficient System Design}
\label{subsection:designConsiderations}
The key enabling concept for energy-neutral indoor localization with the precision of commercial UWB is the shift from an infrastructure-centered to a sensor-centered operation. This allows the initiation of the localization sequence by the mobile tag and, consequently, the power-gating of large circuit parts - including the UWB transceiver - when no localization updates are required. Combined with an accelerometer as a wake-up circuit, this allows an event-driven operation with low-power sleep operation during event-free periods. In an event-driven system, energy consumption is typically dominated by leakage and the consumption of the wake-up circuit. System-level power gating with SiP32431 load switches on every functional block of \cref{fig:blockdiagram} is used to minimize leakage currents. This results in the following main power modes:

\textit{LOCALIZATION} - During a localization event, all power domains are activated with a power consumption of up to \SI{200}{\milli\watt} when UWB messages are received.

\textit{ACTIVE} - In the active power mode, the event and power management circuit is turned on. The 8-bit microcontroller is running at \SI{31}{\kilo\hertz}, and the accelerometer is configured in its lowest power motion wake-up mode. The power gating of the localization domain results in a consumption of \SI{22}{\micro\watt}.

\textit{SLEEP} - By configuring the microcontroller in its sleep mode, the consumption can be further reduced to \SI{4.68}{\micro\watt} while allowing energy harvesting and event-driven motion wake-up functionality.

\textit{DEEP SLEEP} - The potential of the circuit architecture for leakage reduction is recognizable in the lowest power mode of deep sleep. In this configuration,  which is only relevant for the circuit's shelf-life,  the accelerometer and EH circuit are power gated, and only LDO, RTC, and MCU stay supplied. As a wake-up source, the sub-threshold RTC can be used. In order to return to higher power modes, the localization domain has to be activated and the accelerometer reconfigured. The combination of all design steps allows a reduction of circuit leakage to \SI{47}{\nano\ampere}, resulting in a typical consumption of only \SI{173}{\nano\watt}.

\subsection{Event-driven Localization}
\label{subsection:localization}
\begin{figure}[!t]
    \centering
    \includegraphics[width=0.48\textwidth]{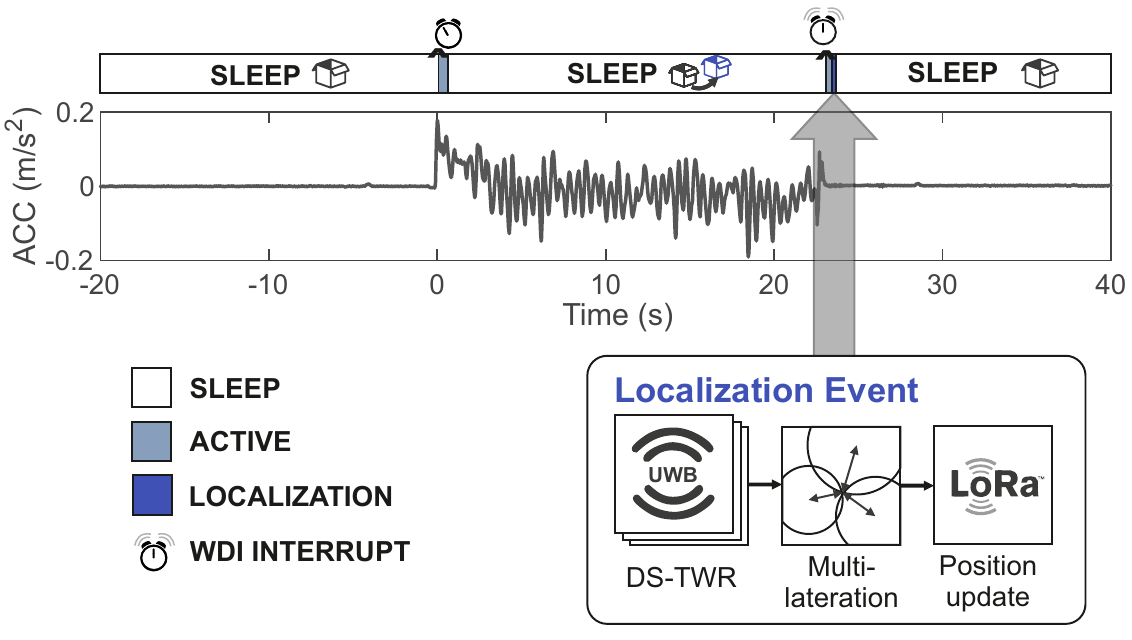}
    \caption{Motion event-driven localization sequence.}
    \label{fig:localizationevent}
\end{figure}

Energy-costly localization events are reduced to a minimum by utilizing the timekeeping features of the energy- and event management circuit and the motion wake-up configuration of the accelerometer. An overview of such an event-driven localization event is given in \cref{fig:localizationevent}. Most of the time, the tag is operating in SLEEP mode with a known position from the previous localization event. If the tag is moved with an acceleration (ACC) higher than the programmed threshold, an interrupt wakes the 8-bit microcontroller of the energy- and event management domain changing the power mode to ACTIVE. The watchdog interrupt (WDI) of the real-time clock is used to detect the end of the motion with minimal energy overhead. By exploiting the recurring interrupts generated by the accelerometer during movements to reset the WDI timer, further interaction with the 8-bit microcontroller is unnecessary, and the circuit can operate in SLEEP mode. Shortly after the motion, a WDI timer overflow will wake the energy- and event management circuit to trigger a new tag localization. To do so, the localization domain is powered up, and the double-sided two-way ranging sequence is started. The tag position is determined with the optimal multilateration algorithm introduced in \ref{sub:trilateration}. Finally, the new position is transmitted with the tag ID to a remote server via LoRa before returning to the initial configuration of SLEEP.

\section{Energy Harvesting \\Characterization and Modeling}
\label{section:results_EH}
To analyze the self-sustainability of the proposed system, the individual functional blocks have been evaluated in terms of functionality, power consumption, and efficiency. This in-depth characterization is subsequently used to build an accurate circuit power model and validate the system's long-time energy neutrality.

The evaluation of the energy consumption during localization events, including a direct comparison between the \textsc{Qorvo} DW1000 and DW3000 UWB transceiver, is given in subsection \ref{subsection:power}. Subsection \ref{subsection:solarCell} shows the characterization of the energy harvesting transducer. Based on the characterization, an accurate power path simulation model is created in subsection \ref{subsection:mbd}. Subsequently, in subsection \ref{subsection:verification}, the model performance is verified in a one-week-long lab experiment in controlled conditions. Finally, subsection \ref{subsection:analyisis} systematical analysis long-time energy neutrality based on simulation.

\subsection{Energy Consumption with Experimental Measurements}
\label{subsection:power}
\begin{figure*}[!t]
    \centering
    \begin{overpic}[width=1\linewidth]{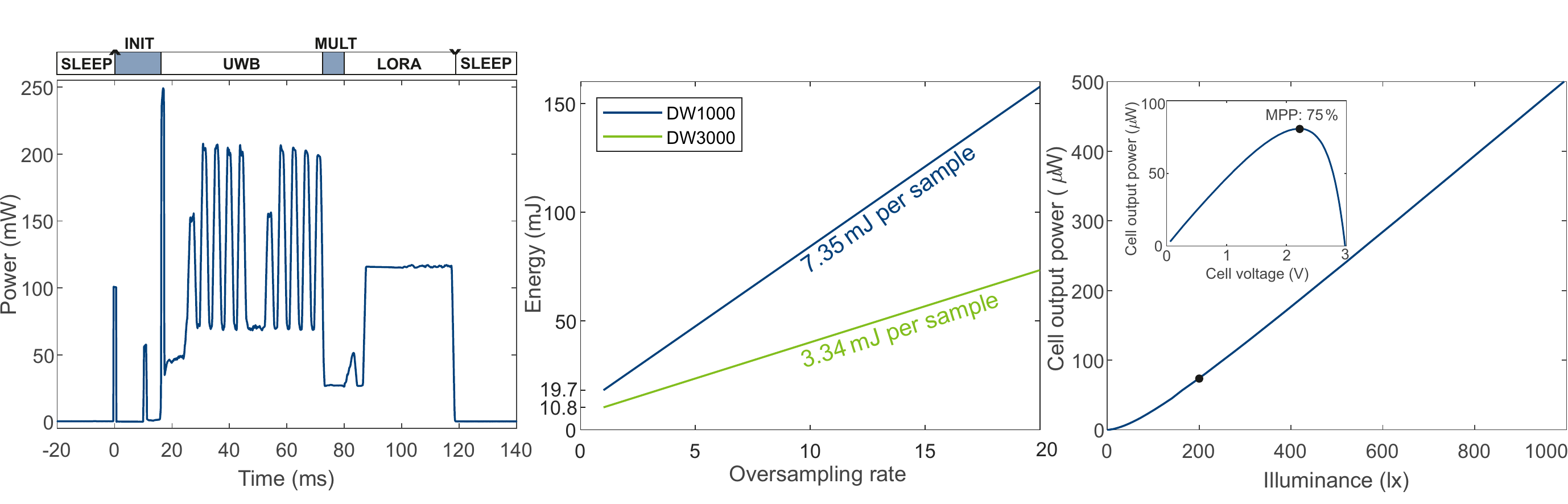}
        \put(0,29){(a)}
        \put(33.33,29){(b)}
        \put(66.66,29){(c)}
    \end{overpic}
    \caption{Circuit power characterization (a) Circuit power consumption during a localization event with deactivated oversampling. (b) Energy consumption per localization event as a function of the oversampling rate.  (c) SP4.2-37 solar cell output power in matched conditions. The inset shows the cell response to a varying load.}
    \label{fig:EH_power}
\end{figure*}

\cref{fig:EH_power} (a) visualizes the power consumption during a localization event with the corresponding states in the top bar. At $t = \SI{0}{\second}$, the localization is started by enabling the related DCDC converters. After \SI{17}{\milli\second}, the 32-bit microcontroller finished the start-up phase and the initialization of the peripherals. The \SI{250}{\milli\watt} inrush spike at \SI{17}{\milli\second} marks the activation of the UWB sub-circuit and the start of the double-sided two-way ranging sequence pictured as one TX peak followed by four RX anchor messages. The sequence is repeated before calculating the tag-anchor distances to allow the two-way error correction based on the algorithm introduced in \ref{sub:doubleSidedTW}. If at least three distances are valid, multilateration is used to calculate the tag position before transmitting the new location to the remote server via LoRa. Finally, after \SI{118}{\milli\second}, the tag returns to its initial SLEEP state. The power consumption has been recorded with a disconnected solar cell and by replacing the battery with a \textsc{Keysight} N6782A source/measurement unit (SMU) module configured to \SI{3.7}{\volt}. A breakdown of the sub-circuits energy consumption for different battery voltages is given in \cref{table:localization}.

\begin{table}[!b]
\caption{Energy consumption during localization event}
\centering
\renewcommand{\arraystretch}{1.35}
\begin{tabular}{@{}l*{3}{c}c@{}}
\toprule
\multirow{2}{*}{Subsection} & \multicolumn{3}{c}{Energy consumption $E_{bat}$} \\
    & $V_{bat} = \SI{3.4}{\volt}$ & $V_{bat} = \SI{3.7}{\volt}$ & $V_{bat} = \SI{4.15}{\volt}$ \\
\midrule
MCU                 & \SI{2.87}{\milli\joule}   & \SI{2.89}{\milli\joule}   & \SI{2.97}{\milli\joule}   \\ 
Multilateration\scriptsize{$^a$}     
                    & \SI{0.23}{\milli\joule}   & \SI{0.24}{\milli\joule}   & \SI{0.31}{\milli\joule}   \\ 
UWB                 & \SI{4.72}{\milli\joule}   & \SI{5.17}{\milli\joule}   & \SI{5.31}{\milli\joule}   \\ 
LoRa                & \SI{2.68}{\milli\joule}   & \SI{2.79}{\milli\joule}   & \SI{2.84}{\milli\joule}   \\ 
\midrule
Total               & \SI{10.28}{\milli\joule}  & \SI{10.84}{\milli\joule}  & \SI{11.13}{\milli\joule}  \\ 
\bottomrule
\end{tabular}
\scriptsize{\\ \vspace{3px} \hspace{150px} $^a$ Part of MCU consumption}\\
\label{table:localization}
\end{table}

The influence of UWB oversampling on the energy consumption per localization event, together with a comparison of the \textsc{Qorvo} DW1000 and DW3000 based UWB module is shown in \cref{fig:EH_power} (b). The novel DW3000 model reduces the energy per ranging event by \SI{45}{\percent} in a similar configuration. The energy cost for oversampling is \SI{7.35}{\milli\joule} and \SI{3.34}{\milli\joule} for the DW1000 and DW3000 IC, respectively. Similarly to the oversampling, the node's energy consumption will scale linearly with the number of event-driven activations.  


\subsection{Characterization Energy Harvesting Transducer} 
\label{subsection:solarCell}
To optimally match the solar cell with the energy harvesting circuit, the CP4.2-37 cell was first characterized in open circuit and matched conditions. \cref{fig:EH_power} (c) shows the maximal output power of the cell in indoor lighting conditions. For the measurement, the cell was placed in a darkened chamber artificially illuminated with a broadband light source while the output was monitored and loaded with a \textsc{Keysight} B2902A SMU. The inset shows the solar cell's typical response to a varying load at a constant illuminance of \SI{200}{\lux} and a maximal power point at \SI{75}{\percent} of the open-circuit voltage.

\subsection{Circuit Model Creating} 
\label{subsection:mbd}
Analyzing energy neutrality over the targeted years of operating time inevitably requires an analytical approach. For that reason, the information collected on the energy harvesting transducer and the circuit is used to create an accurate system power path model used as a foundation for an in-depth analysis. Applying a model-based design approach in combination with indoor light datasets allows not just the simulation of harvested and consumed energy but also the validation of design decisions, such as the energy storage element selection \cite{Mayer-MBD}.

\vspace{2mm}\subsubsection{Indoor lighting dataset} The indoor solar harvesting dataset presented in \cite{gomezACM} is used to analyze self-sustainability in-depth. In version 2.0, it contains, among other data, the illuminance logged in an office environment over approximately two years. In particular, the dataset comprises two low-light office locations with little natural light (P06 and P13) and two office locations with significant natural light and partly direct sunlight (P14 and P17). In addition to that, the dataset contains with location P18, a very-low light measurement that can be used as a baseline for the system’s shelf-life. An overview of the positions and the associated illuminance during the day, together with the standard deviation, is given in \cref{table:dataset}. The daily energy column exemplifies a thin-film solar cell's energy harvesting potential per square centimeter.

\begin{table}[!b]
\caption{Dataset Solar Energy Harvesting Potential}
\centering
\renewcommand{\arraystretch}{1.35}
\begin{tabular}{@{}l*{2}{c}c@{}}
\toprule
Position  & \hspace{15px} Illuminance (day) \hspace{15px}  & \hspace{15px} Daily Energy \hspace{15px}    \\
\midrule
P06     & 112 $\pm$ \SI{52}{\lux}   & 54 $\pm$ \SI[per-mode = symbol]{33}{\milli\joule \per\centi\metre\squared}   \\  
P13     & 101 $\pm$ \SI{46}{\lux}   & 49 $\pm$ \SI[per-mode = symbol]{32}{\milli\joule \per\centi\metre\squared}   \\
P14     & 464 $\pm$ \SI{313}{\lux}  & 393 $\pm$ \SI[per-mode = symbol]{331}{\milli\joule \per\centi\metre\squared} \\
P17     & 356 $\pm$ \SI{172}{\lux}  & 294 $\pm$ \SI[per-mode = symbol]{191}{\milli\joule \per\centi\metre\squared} \\
P18     & 28 $\pm$ \SI{3}{\lux}     & 11 $\pm$ \SI[per-mode = symbol]{7}{\milli\joule \per\centi\metre\squared }    \\
\bottomrule
\end{tabular}
\label{table:dataset}
\end{table}

\vspace{2mm}\subsubsection{Circuit simulation model} 
\begin{figure}[!t]
    \centering
    \includegraphics[width=0.50\textwidth]{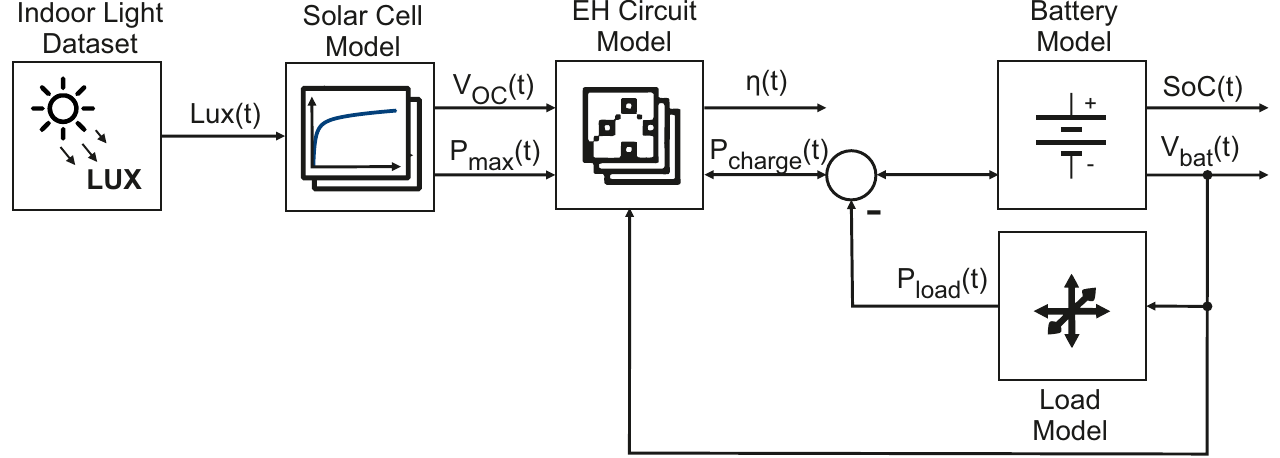}
    \caption{The high-level overview of the circuit's power model.}
    \label{fig:simulation_blockdiagram}
\end{figure}

A comprehensive simulation model has been derived based on exhaustive lab measurements to allow a realistic evaluation of long-term self-sustainability. The high-level overview of the sensor node's power path implemented as a multi-domain \textsc{Matlab Simulink} simulation model is shown in \cref{fig:simulation_blockdiagram}.

\textit{Solar Cell Model} - The solar cell behavior in matched conditions can be directly described with higher-order polynomials based on the characterization subsection \ref{subsection:solarCell}.

\textit{EH Circuit Model} - To fully characterize the circuit's energy conversion efficiency, the harvesting circuit was tested in all relevant operational states by replacing the energy-harvesting transducer and battery with a \textsc{Keysight} B2902A SMU. Subsequently, a Random Forest regression model was trained to calculate the circuit efficiency as a function of input voltage, input power, and battery voltage. In addition to the energy conversion behavior, the circuit model considers the consumption of the energy and power management domain and the overcharge protection of the energy storage element.

\textit{Load Model} - Due to the power gating of the localization circuit, the host system is typically modeled solely by consideration of the accelerometer consumption in wake-up mode. The event-driven activation of the localization circuit is accounted for, based on the circuit's energy consumption measured in subsection  \ref{subsection:power}.

\textit{Battery Model} - The characteristics of the battery storage element can be directly emulated by a \textsc{Matlab Simulink} generic battery model block. To also model the battery self-discharge, the leakage is considered LiPo typical as a percentage of the state of charge (SoC) per month.

\vspace{2mm}

\subsection{Circuit Model Verification}
\label{subsection:verification}
The harvesting circuit of the physical prototype is analyzed under controllable and repeatable environmental conditions to verify the created system model. This is achieved by placing the solar cell in a darkened chamber, artificially illuminated by a controllable broadband light source. Thus, the circuit behavior can be precisely monitored while emulating realistic indoor light scenes. In particular, the following experimental setup has been used to determine unknown battery parameters and assess the simulation model’s performance:

The illuminance is controlled and monitored with two software-controlled source/measurement units, \textsc{Keysight} N6782A, connected to a halogen lamp and a photoresistor. Currents and voltages are logged with \textsc{Keysight} 34465A multimeters in high-z input configuration between solar cell and energy harvesting converter as well as between converter and battery. Finally, the load is emulated with a highly precise \textsc{Keysight} B2902A SMU connected parallel to the battery.

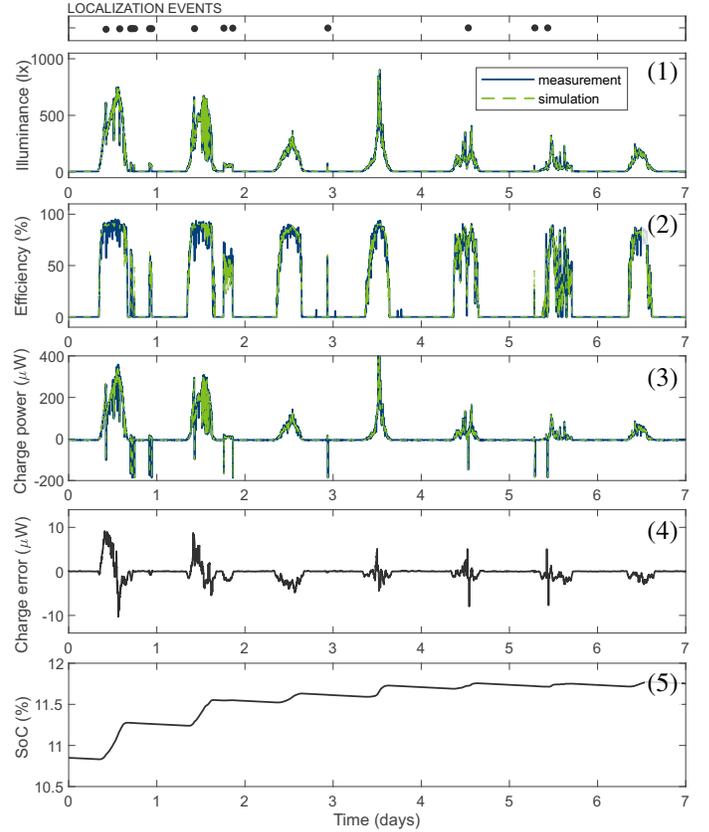
\begin{figure}[!t]
    \centering
    \begin{overpic}[width=0.5\textwidth]{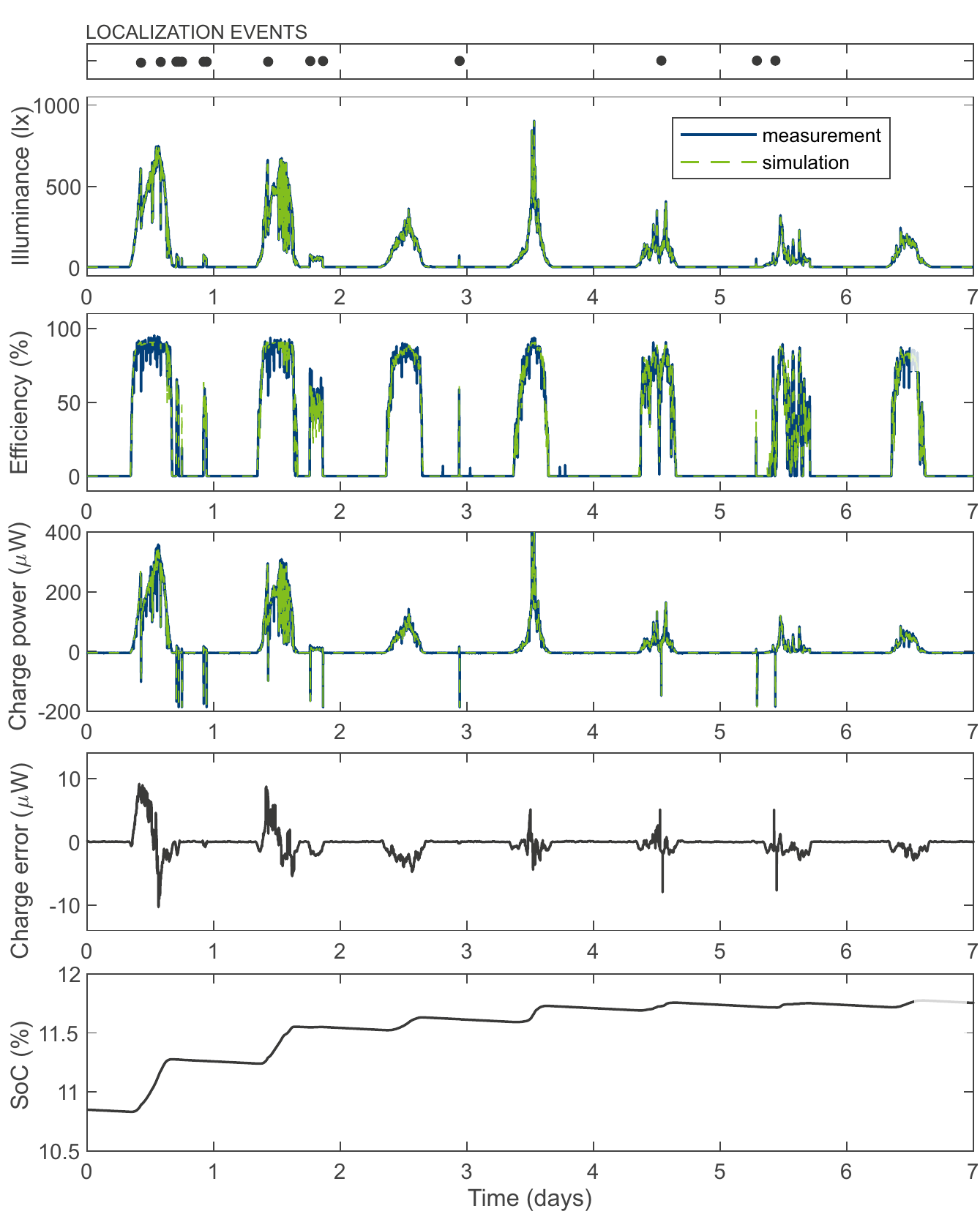}
        \put(75,17.2){(5)}
        \put(75,35){(4)}
        \put(75,53){(3)}
        \put(75,71){(2)}
        \put(75,89){(1)}
    \end{overpic}
    \caption{Model verification based on a one-week-long measurement under controlled lab conditions.}
    \label{fig:verification}
\end{figure}

\cref{fig:verification} shows the model verification results from a one-week-long controlled lab experiment reproducing the illuminance of dataset \cite{gomezACM} on position P17. Errors caused by degeneration of the lamp during data acquisition are compensated by using the measured illuminance as an input source for the simulation model. For the measurement, the event-driven system behavior is mimicked with random localization events visualized in the top bar of the figure. The first subplot of \cref{fig:verification} illustrates the illuminance in lux emulating a dynamic indoor light scene with partially direct sunlight. The second subplot shows the conversion efficiency of the EH boost converter reaching up to \SI{91}{\percent} during bright scenes. Subplots three and four show the battery charge power and the corresponding error expressed in \si{\micro\watt} between measurement and simulation. During the night or localization events, the system operates predominantly from the energy stored in the battery visualized by a negative charge current. The error shown in subplot four is visualized as a mean calculated of a sliding window with a length of \SI{15}{\minute}. Finally, the lowest subplot illustrates the predicted battery state-of-charge (SoC) changes due to the harvested energy.

The model performance metrics resulting from the verification experiment are summarized in \cref{table:results_lipo_model}, where $\hat{P}_{CH}$ represents the predicted battery charge power of the model and $P_{CH}$ the measured counterpart. The results show a root mean square error (RMSE) of \SI[per-mode = symbol]{6.3}{\micro\watt} for the power balance between measurement and simulation in the \SI{7}{\day} long period. The model's coefficient of determination can be calculated according to \cref{eq:r2} as 0.988. By integrating the measured and simulated battery charge power over time, the energy error $(E_{err\,\%})$ can be calculated following \cref{eq:energyError} as \SI{-1.18}{\percent} in the analyzed timeframe. The precise replication of the circuit dynamics demonstrates simulation accuracy with a negligible error compared to component tolerances, thus allowing a reliable analysis of the system's long-time behavior.

\begin{table}[!b]
\caption{Model verification results}
\centering

\resizebox{0.5\textwidth}{!}{%
\renewcommand{\arraystretch}{1}
\begin{tabular}{@{}l*{1}{c}c@{}}
\toprule
Metric    & Definition & Result \\
\midrule
$RMSE\small$ & \begin{minipage}{5cm}{\begin{equation} \sqrt{\frac{1}{n}\sum_{i=1}^{n} \left( \hat{P}_{CH,i}-P_{CH,i} \right)^2 } \end{equation}}\end{minipage} & \SI{6.3}{\mu\watt}\\

$R^2$ &\begin{minipage}{5cm}{\begin{equation} 1-\frac{\sum_{i=1}^{n} \left( \hat{P}_{CH,i}-P_{CH,i} \right)^2 }{\sum_{i=1}^{n} \left( P_{CH,i}-\frac{1}{n}\sum_{j=1}^{n}  P_{CH,j} \right)^2} \label{eq:r2} \end{equation}}\end{minipage} & 0.988\\

$E_{err\,\%}$ & \begin{minipage}{5cm}{\begin{equation} \frac{\int_{}^{}{\hat{P}_{CH,i}(t)\,dt}}{\int_{}^{}{P_{CH,i}(t)\,dt}}-1 \label{eq:energyError} \end{equation}} \end{minipage}& \SI{-1.18}{\percent}\\
\bottomrule
\end{tabular}
}
\label{table:results_lipo_model}
\end{table}

\subsection{Analysis Long-time Energy Neutrality}
\label{subsection:analyisis}
\begin{figure*}[!t]
    \centering
    \begin{overpic}[width=1\linewidth]{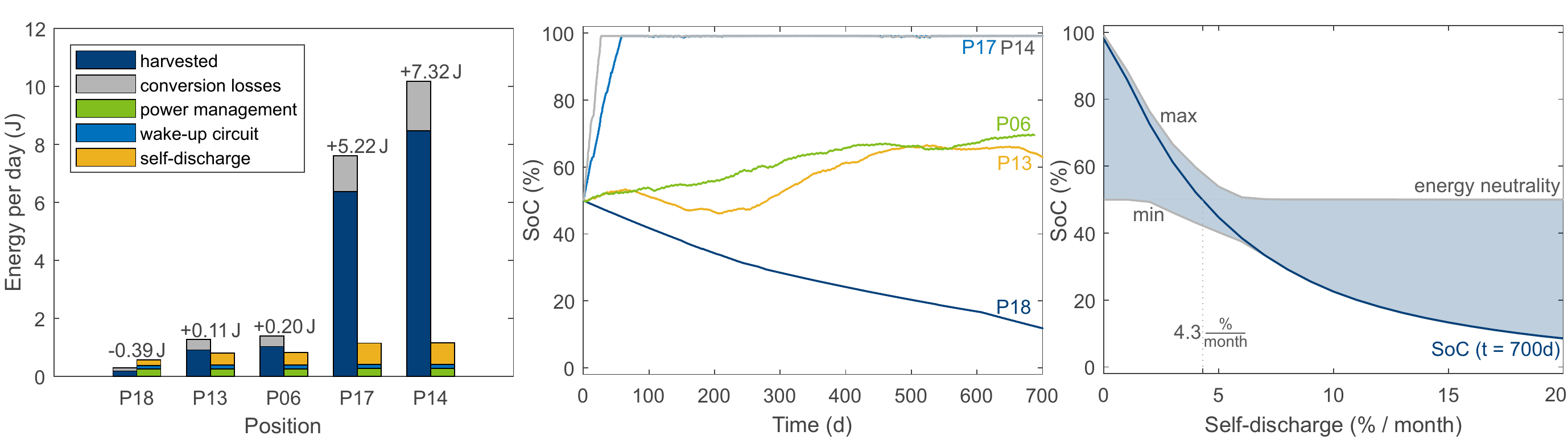}
        \put(0,28){(a)}
        \put(33.33,28){(b)}
        \put(66.66,28){(c)}
    \end{overpic}
    \caption{Model-based design long-time evaluation of the proposed self-sustainable indoor localization circuit. (a) Average energy balance in five different indoor lighting conditions. (b) Battery state of charge in idle condition. (c) Influence of the battery self-discharge on the energy balance.}
    \label{fig:simulation}
\end{figure*}

The results of the long-time energy-neutrality analysis are summarized in \cref{fig:simulation}. Subfigure (a) shows the mean harvested and consumed energy per day for a tag deployed at the five previously mentioned positions. 
A battery self-discharge rate of \SI[per-mode = symbol]{3}{\percent\per month} is considered for the simulation, and the overcharge protection is configured to \SI{4.2}{\volt}. To decouple the simulation from a specific load profile caused by localization events, the energy surplus is considered to be consumed by the circuit. Thus, the overcharge protection is assumed not to influence the harvested energy and the corresponding harvesting losses. The initial state of charge has been configured to \SI{50}{\percent}.

Due to the underlying dataset's high illuminance variability, the harvested energy per day varies significantly. On position P18, the circuit shows a negative energy balance despite being able to harvest \SI[per-mode = symbol]{0.19}{\joule\per\day} on average. In the low-light locations, P13 and P06, the circuit achieves on average an energy surplus of \SI[per-mode = symbol]{0.11}{\joule\per\day} and \SI[per-mode = symbol]{0.20}{\joule\per\day}, respectively. The surplus allows to perform 10 to 18 localizations per day or to charge the battery further. Significant more energy can be harvested at the office locations with natural light (P17 and P14). The typical \SI[per-mode = symbol]{5.2}{\joule\per\day} and \SI[per-mode = symbol]{7.3}{\joule\per\day} result in a fully charged battery or 480 and 670 localizations per day. Vital for the application, despite a small gap between harvested power and idle consumption, is the capacity of the energy storage element. The \SI{50}{\milli\ampere\hour} battery allows energy storage for up to \SI{62}{k} localization events, ensuring operation also during energy-dry periods. 

A closer look at the consumed energy clearly shows that the battery self-discharge dominates the system's idle consumption, especially when the battery gets fully charged, as shown in \cref{fig:simulation} (b). In the case of P18, the negative energy balance results in a battery discharge from \SI{50}{\percent} to \SI{11}{\percent} in the analyzed \SI{700}{\day} long timeframe. In contrast to that, the positions with natural light charge the battery fully in less than two months. P06 and P13 illustrate the importance of the long-time analysis with energy-positive and negative phases depending on the season. 

Finally, \cref{fig:simulation} (c) analysis the influence of the battery self-discharge on the energy balance, which is a critical factor of efficiently designed sensor nodes \cite{Yue2020}. For the simulation, the illuminance of position P13 has been chosen as the input source, and the self-discharge rate is varied. The curve shows the state of charge after \SI{700}{\day} with minimum and maximum values as a shaded envelope during this period. Up to a self-discharge rate of \SI[per-mode = symbol]{4.3}{\percent\per month}, the system reaches energy neutrality with a  state of charge of \SI{50}{\percent} at the beginning and end of the time period.

\section{Indoor Localization \\ Application-specific Evaluation}
\label{section:results_localization}
To evaluate the localization performance of the designed tag depicted in \cref{fig:randering}, localization accuracy is first analyzed in 2D and 3D. Subsequently, the system's indoor positioning capability is used for machine learning-based asset tracking in a real-world office environment.

The single- and multi-room localization accuracy is validated in subsection \ref{subsection:accuracy}. Subsections \ref{subsection:assetTracking} and \ref{subsection:discussion} conclude the evaluation by demonstrating the accuracy during indoor asset tracking and discussing the limitations of the design.

\subsection{Localization Accuracy in Real-world Environment}
\label{subsection:accuracy}
\begin{figure*}[!t]
    \centering
    \begin{overpic}[width=1\linewidth]{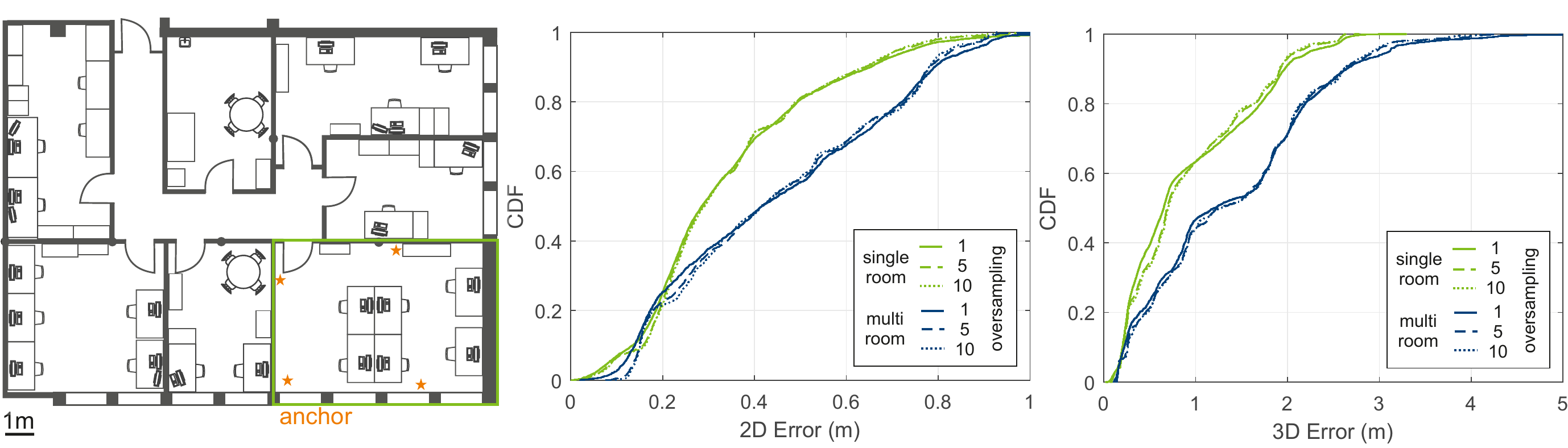}
        \put(0,28){(a)}
        \put(33.33,28){(b)}
        \put(66.66,28){(c)}
    \end{overpic}
    \caption{Localization accuracy in single- and multi-room environments. (a) Floor plan of the \SI{200}{\square\meter} office space with marked UWB anchors. (b) CDF of the 2D error calculated as Euclidean distance. (c) CDF of the 3D error.}
    \label{fig:CDF}
\end{figure*}

The ranging accuracy has been analyzed independently of the energy consumption in a multi-room environment. The experimental setup, shown in \cref{fig:CDF} (a), is an over \SI{200}{\square\meter} office space composed of seven rooms. The environment has been selected purposely non-idealized, with partly steel-reinforced concrete walls, working staff, and multiple metal shelves and cabinets. The UWB anchors, visualized as yellow stars in \cref{fig:CDF}, have been mounted on tripods at the height of \SI{2}{\meter} and \SI{1.5}{\meter}. The data has been acquired based on two individual datasets (single-room: 51 positions, 6k samples, multiroom: 18 positions, 2k samples). Employees regularly used the office space during data acquisition, and the node antenna was rotated at individual testing locations. All data processing was conducted in real-time on the sensor node running the embedded algorithms presented in section \ref{section:background}. The ground truth was determined with a laser distance meter to calculate the absolute error.

\cref{fig:CDF} (b) and (c) show the cumulative distribution function (CDF) of the localization error when applying a 2D and 3D  multilateration, respectively. The results suggest that oversampling and averaging do not positively impact the localization error. This is due to the system’s high precision but limited accuracy and is caused by the antenna orientations and environmental influences, such as walls and shelves, in the realistic setup. The system executing the trilateration algorithm introduced in \ref{sub:trilateration} reaches an average error of \SI{28}{\centi\meter} and \SI{50}{\centi\meter} in the single-room setup for 2D and 3D localization. For all rooms, an error of \SI{0.4}{\meter} in 2D and \SI{1.2}{\meter} in 3D was reached. It is important to point out that overall accuracy in the multi-room scenario increases due to the data points collected in the proximity of the anchors. In the acquired multi-room dataset, \SI{32}{\percent} of the samples are collected in the room with anchors, increasing the mean measured localization accuracy by \SI{8}{\centi\meter} (2D) and \SI{17}{\centi\meter} (3D).
The room layout combined with the anchor placement also showcases the limitations of the UWB-based system. By leaving the office space via the north exit, signal reception is too low for reliable double-sided two-way ranging. Aside from that, the anchor placement in the corner of the office space implies that up to three times the area could be covered with a four-anchor system and an optimal floor plan.

\subsection{Indoor Asset Tracking}
\label{subsection:assetTracking}
\begin{figure*}[!t]
    \centering
    \begin{overpic}[width=1\linewidth]{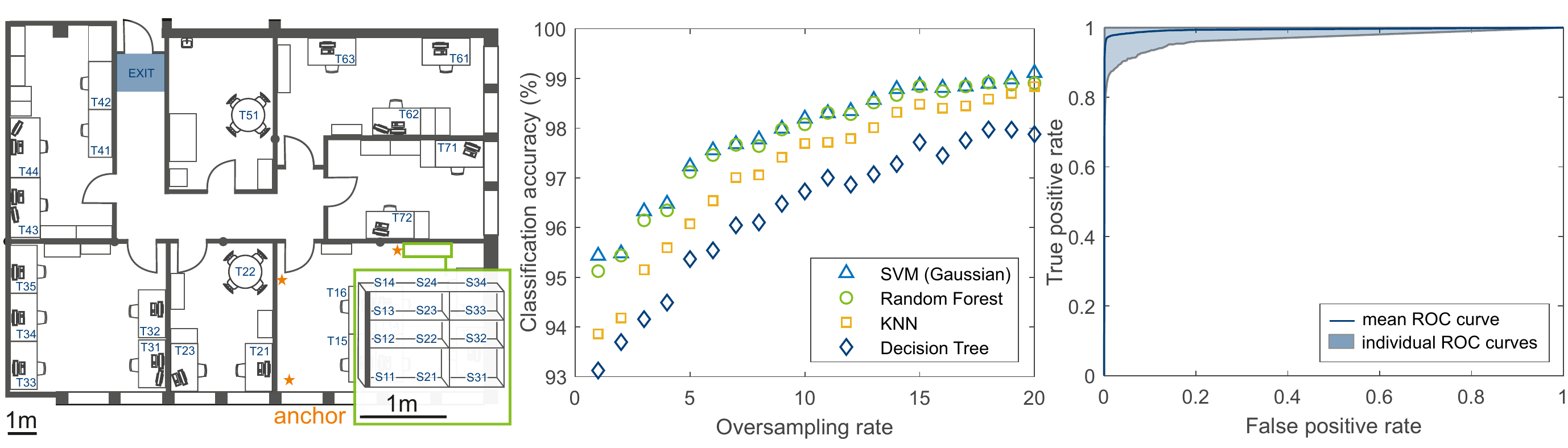}
        \put(0,28){(a)}
        \put(33.33,28){(b)}
        \put(66.66,28){(c)}
    \end{overpic}
    \caption{Indoor asset tracking application based on the proposed self-sustainable sensor. (a) Floor plan with marked location classes. The inset shows the shelf positioned in the room with the anchors to achieve maximum classification accuracy. (b) Classification accuracy of the trained classifiers as a function of the oversampling rate. (c) Mean ROC curve of a random forest classifier trained on data with deactivated oversampling.}
    \label{fig:localization}
\end{figure*}

Beyond the pure localization error, the system has been analyzed in the targeted application of indoor asset tracking. For this purpose, an independent dataset has been acquired containing 24 desk- and 12 shelf-positions visualized in \cref{fig:localization} (a). In addition to fixed positions, data was collected while carrying the tag in the office space and close to the exit. As  input features, the on-board processed three-dimensional multilateration output was used. The tag was mounted on the exterior of a plastic box during the data acquisition, and care was taken to acquire data from the full table surface. Overall this results in 38 labeled classes, each represented by more than 500 data points.

\cref{fig:localization} (b) shows the comparison of the classification accuracy, defined as the sum of the correctly predicted labels (True Positive, False Positive) over all classified samples (True Positive, True Negative, False Positive, False Negative) according to \cref{equation:accuracy}, for different classifiers.
\begin{equation}
Accuracy = \frac{TP+TN}{TP+TN+FP+FN}
\label{equation:accuracy}
\end{equation}
To avoid overfitting, 10-fold cross-validation is used. A support vector machine (SVM) with a Gaussian kernel and a random forest-based classifier allows distinguishing between the different locations with a classification accuracy of over \SI{95}{\percent} without UWB oversampling and averaging. Despite the negligible influence on the absolute error shown in the previous section, oversampling reduces outliers and thus enhances the classification accuracy to \SI{99}{\percent} for rates higher than 14. \cref{fig:localization} (c) visualizes the accompanying receiver operating characteristic (ROC) curve for the random forest classifier and an oversampling rate of one.

The splitting of the dataset into subsets for table- and shelf-positions, summarized in \cref{table:accuracySplit}, details the system's limitations. Misclassifications occur predominantly between the individual shelf positions due to the importance of height information. In addition to the height limitations, the \textit{non-classified} class, collected while carrying the tag, comprises hard-to-classify outliers when oversampling is deactivated, limiting the accuracy. In contrast to that, table positions can be detected accurately in the entire office space, resulting in a comparable localization performance for the complete dataset (multi-room) and its single-room subset.
\begin{table}[!b]
\caption{Random forest classification accuracy for different data subsets as a function of the oversampling rate.}
\centering
\renewcommand{\arraystretch}{1.35}
\begin{tabular}{@{}ll*{4}{c}c@{}}
\toprule
& \multirow{2}{*}{Subset} \hspace{15px}  & \multicolumn{4}{c}{\phantom{.}\hspace{50px}Oversampling\hspace{50px}\phantom{.}} \\
      &  & 1 & 5 & 20 & 15\\
\midrule
\rule{0px}{0px} \parbox[t]{2mm}{\multirow{4}{*}{\rotatebox[origin=c]{90}{accuracy}}} & table           & \SI{97.93}{\percent}   & \SI{98.77}{\percent}  & \SI{99.27}{\percent}   & \SI{99.48}{\percent}\\  
&shelf           & \SI{91.49}{\percent}   & \SI{94.50}{\percent}  & \SI{96.99}{\percent}   & \SI{98.21}{\percent}\\  
&singe-room      & \SI{95.18}{\percent}   & \SI{96.82}{\percent}  & \SI{98.14}{\percent}   & \SI{98.90}{\percent}\\  
&multi-room      & \SI{95.19}{\percent}   & \SI{97.06}{\percent}  & \SI{98.20}{\percent}   & \SI{98.85}{\percent}\\  
\bottomrule
\end{tabular}

\label{table:accuracySplit}
\end{table}

The measurement results clearly show the system's applicability for the asset tracking scenario. The localization accuracy limiting factors, such as anchor orientation and static environmental influences, can be compensated sufficiently during the training phase.

\subsection{Discussion}
\label{subsection:discussion}
This work is motivated by the vision of an unobtrusive and highly accurate indoor asset tracking system with conservative infrastructure overhead. It is evident that the opposing design objectives of unobtrusiveness, and therefore the requirement for energy autonomy, on the one hand, and the high accuracy, on the other hand, are not without any limitations. 

Opting for state-of-the-art commercial UWB transceivers combined with double-sided two-way ranging has the advantage of reaching a few tens of meters communication range and robust localization even in realistic NLOS environments. These advantages come at the cost of significantly increased energy consumption during localization. By utilizing event-driven sensing based on the tag-centralized approach, the impact of the localization system on the overall energy budget can be reduced significantly. However, energy savings based on event-driven operation strongly depend on the incidence of localization events. If applications with a high localization sampling rate, such as trajectory tracking, are targeted, the low environmental energy will be insufficient to serve the circuit's energy demand. Thus the presented work is limited to applications that tolerate sparse location updates.

In addition, despite reaching over \SI{100}{\metre} in LOS scenarios, the building structure inevitably highly affects UWB-based systems applied indoors. Thus, for large-scale settings, a denser anchor grid is necessary. This is particularly relevant if 3D localization accuracies below a half meter are required. In the presented example of subsection \ref{subsection:assetTracking}, shelf positions could only be reliably determined due to the anchor placement in the proximity of the shelf.


\section{Conclusion}
\label{section:conclusion}
This paper presented the design of an energy-neutral indoor asset tracking system. This includes a hardware-software co-designed event-driven tag and full characterization of the proposed circuits and algorithms and their influence on the system's energy budget. The heterogeneous circuit has been designed for nano-quiescent current operation, by offloading control tasks to a programmable, highly energy-efficient energy- and event-management domain. Combined with an accelerometer in motion wake-up configuration, this allows an event-driven activation of the ultra-wideband-based localization circuit. The experimental results demonstrate the benefits of the design with a quiescent current of \SI{47}{\nano\ampere} in the most efficient power mode of deep-sleep and \SI{1.2}{\micro\ampere} with active energy harvesting and motion wake-up circuit.

The implemented double-sided two-way ranging combined with the embedded optimal multilateration reaches an average localization error of \SI{28}{\centi\meter} and \SI{50}{\centi\meter} in the single-room office setup for 2D and 3D localization, respectively. In the multi-room setup, \SI{0.4}{\meter} in 2D and \SI{1.2}{\meter} in 3D were reached. These results were achieved in a realistic and dynamic office environment with four anchor nodes and without calibration or optimal alignment of antennas. Applied for asset tracking in a \SI{200}{\metre\squared} office space, an object with an attached sensor node can be localized with an accuracy of over \SI{95}{\percent}  (38 classes). 

Finally, the system's long-time self-sustainability has been analyzed over \SI{700}{\day} in multiple indoor lighting situations. The \SI{50}{\milli\ampere\hour} sized battery enables up to \SI{62}{k} localization with an energy consumption per localization of \SI{10.84}{\milli\joule}. The energy harvesting subsystem allows next to the balancing of the circuit's idle consumption and the battery self-discharge for an additional 10 to 500 localization per day in typical indoor lighting conditions.

\section*{Acknowledgments}
The authors would like to thank Thiemo Zaugg, Tobias Margiani, and Christoph Schnetzler for their commitment during their bachelor's and master's projects.

\bibliography{IEEEabrv,./main}

\begin{IEEEbiography}[{\includegraphics[width=1in,height=1.25in,clip,keepaspectratio]{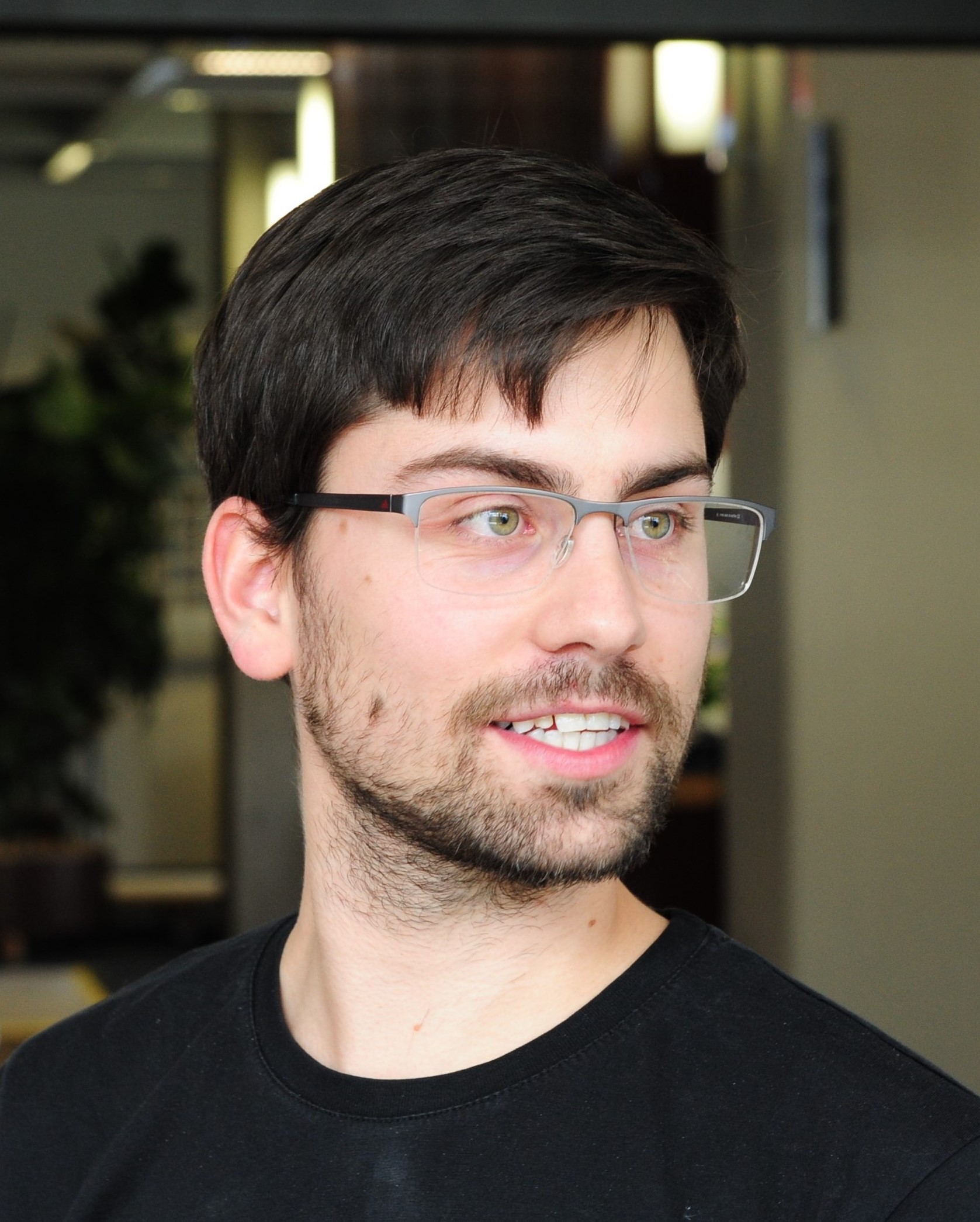}}]{Philipp Mayer}
 (Graduate Student Member, IEEE) received the B.Sc. degree in electrical engineering and information technology from TU Wien, Vienna, Austria, in 2016, and the M.Sc. degree from ETH Zurich, Zurich, Switzerland, in 2018, where he is currently pursuing the Ph.D. degree with the Integrated System Laboratory. \\
His research interests include low-power system design, energy harvesting, and edge computing. \\
Mr. Mayer was a recipient of the Best Paper Award at the 2017 IEEE International Workshop on Advances in Sensors and Interfaces and the Best Student Paper Award at the 2018 IEEE Sensors Applications Symposium. Beyond his particular area of expertise, he was granted the Best Poster Award in the 2018 IOP Workshop in Devices, Materials and Structures for Energy Harvesting and Storage. In 2019, he founded Mayer Engineering and Consulting, intending to connect academics with industrial expertise.
\end{IEEEbiography}

\begin{IEEEbiography}[{\includegraphics[width=1in,height=1.25in,clip,keepaspectratio]{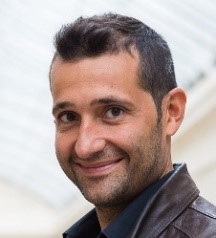}}]{Michele Magno}
(Senior Member, IEEE) received he master's and Ph.D. degrees in electronic engineering from the University of Bologna, Bologna, Italy, in 2004 and 2010, respectively. \\
Currently, he is a Senior Researcher at ETH Zurich, Zurich, Switzerland, where he is the Head of the Project-Based Learning Center. He has collaborated with several universities and research centers, such as Mid University Sweden, where he is a Guest Full Professor. He has published more than 150 articles in international journals and conferences, in which he got multiple best paper and best poster awards. The key topics of his research are wireless sensor networks, wearable devices, machine learning at the edge, energy harvesting, power management techniques, and extended lifetime of battery-operated devices.
\end{IEEEbiography}

\begin{IEEEbiography}[{\includegraphics[width=1in,height=1.25in,clip,keepaspectratio]{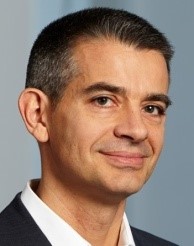}}]{Luca Benini}
(Fellow, IEEE) received the Ph.D. degree in electrical engineering from Stanford University, Stanford, CA, USA, in 1997.\\
He has served as the Chief Architect of the Platform2012/STHORM Project with STMicroelectronics, Grenoble, France, from 2009 to 2013. Currently, he holds the Chair of Digital Circuits and Systems at ETH Zurich, Zurich, Switzerland, and is a Full Professor at the University of Bologna, Bologna, Italy. He has published more than 1000 peer-reviewed articles and five books. His current research interest includes energy-efficient computing systems' design from embedded to high performance.\\
Dr. Benini is a fellow of the ACM and a member of the Academia Europaea. He was a recipient of the 2016 IEEE CAS Mac Van Valkenburg Award and the 2019 IEEE TCAD Donald O. Pederson Best Article Award.
\end{IEEEbiography}

\vfill

\end{document}